\documentclass[
superscriptaddress,
nofootinbib,
 amsmath,amssymb,
 aps,
pra,
twocolumn]{revtex4-2}

\usepackage[utf8]{inputenc}
\usepackage{graphicx}
\usepackage{hyperref}
\hypersetup{colorlinks = true, linkcolor = [rgb]{0.19411,0.51882,0.667058}, urlcolor = [rgb]{0.125490,0.29542,0.1647058}, citecolor = [rgb]{0.75882,0.37411,0.14117}, breaklinks=true}
\usepackage{amsmath}
\usepackage{amsfonts}
\usepackage{amssymb}
\usepackage{mathtools}
\usepackage{dsfont}
\usepackage{braket}
\usepackage{comment}
\usepackage[format=plain]{caption}
\usepackage{subcaption}
\newcommand*{\I}{\mathrm{i}\mkern1mu}
\newcommand{\sep}{\overset{\text{sep.}}{\geq}}

\DeclarePairedDelimiter\abs{\lvert}{\rvert}

\begin{document}

\title{Revealing entanglement through local features of phase-space distributions}

\author{Elena Callus}\email{elena.callus@uni-jena.de}
\affiliation{Institute of Condensed Matter Theory and Optics, Friedrich-Schiller-Universit\"{a}t Jena, Max-Wien-Platz 1, 07743 Jena, Germany}
\author{Martin G\"{a}rttner}\email{martin.gaerttner@uni-jena.de}
\affiliation{Institute of Condensed Matter Theory and Optics, Friedrich-Schiller-Universit\"{a}t Jena, Max-Wien-Platz 1, 07743 Jena, Germany}
\author{Tobias Haas}\email{tobias.haas@uni-ulm.de}
\affiliation{Institut für Theoretische Physik and IQST, Universität Ulm, Albert-Einstein-Allee 11, 89069 Ulm, Germany}

\begin{abstract}
We formulate an infinite hierarchy of continuous-variable separability criteria in terms of quasiprobability distributions and their derivatives evaluated at individual points in phase space. Our approach is equivalent to the Peres--Horodecki criterion and sheds light on how distillable entanglement manifests in the phase-space picture. We demonstrate that already the lowest-order variant constitutes a powerful method for detecting the elusive non-Gaussian entanglement of relevant state families. Further, we devise a simple measurement scheme that relies solely on passive linear transformations and coherent ancillas. By strategically probing specific phase-space regions, our method offers clear advantages over existing techniques that rely on access to the full phase-space distributions.

\end{abstract}

\date{\today}

\maketitle

\section{Introduction}

Entanglement is a central resource for quantum information processing~\cite{Ekert1998,Bergou2005}, quantum computing~\cite{Jozsa1998,Brassard1998,Preskill2012}, quantum communication~\cite{Gisin2007} and metrology~\cite{Giovannetti2011,Huang2024}. Together with a fundamental interest in understanding this genuinely quantum phenomenon~\cite{Wootters1998,Horodecki2009}, this renders the efficient certification of entanglement an essential problem. Entanglement detection becomes particularly challenging in the continuous-variable (CV) paradigm, due to the infinite-dimensional Hilbert space. In particular, there is a need for robust entanglement witnesses accessible within current experimental capabilities.

CV systems encode information in continuous eigenspectra and serve as prominent platforms for various quantum information processing tasks~\cite{Braunstein2005,Weedbrook2012,Serafini2017,Slussarenko2019}. Physical implementations include quantum-optical setups~\cite{Schleich2001,Klauder2006,Mandel2013}, where the canonically conjugate quadratures of the electromagnetic field act as position and momentum observables, as well as ion-trap and ultracold-atom systems~\cite{Pitaevskii2016}. Although entanglement theory for continuous variables is less developed compared to the discrete-variable case, CV systems have recently been growing in relevance and are gaining more interest for computing~\cite{Kok2007,Buck2021,Hahn2025,Romero2025,AghaeeRad2025} and simulation purposes~\cite{AspuruGuzik2012,Dachille2025}.

Several approaches have been adopted to construct entanglement witnesses in the CV realm. This includes criteria based on moments of the field mode operators~\cite{Duan2000,Simon2000,Mancini2002,Giovannetti2003,Shchukin2005,Agarwal2005,Griffet2023b,Callus2025} and of the partially transposed state~\cite{Deside2025}, entropic witnesses~\cite{Walborn2009,Walborn2011,Saboia2011,Tasca2013,Schneeloch2018,Haas2021b,Haas2022a} and conditions based on continuous majorization theory~\cite{Haas2023a,Haas2023b}. However, a key drawback common to many of these methods is that estimating relevant observables can become very resource-intensive, especially for states that are notoriously hard to witness. Strongly non-Gaussian states often require estimating moments beyond second order, which rapidly becomes expensive in terms of sampling complexity. Similarly, entropic criteria require estimating (quasi-)probability distributions, which is very expensive. Additionally, experimental efforts are shifting away from utilising homodyne detection towards photon-number-resolving detection. The latter remains in the process of rapid developments~\cite{Kuhn2017,Wang2017,Weinbrenner2024,Cheng2022,Wang2025}, motivated by its necessity in applications such as boson sampling~\cite{Aaronson2013,Hamilton2017,Madsen2022,Dellios2025,Liu2025}, preparation of non-Gaussian states~\cite{Eaton2019,Walschaers2021,Eaton2022}, as well as multiphoton metrology~\cite{Slussarenko2017}.

In our work, we formulate entanglement witnesses in terms of $\sigma$-parametrized phase-space distributions~\cite{Cahill1969,Cahill1969_2} at arbitrary points in phase space. This complements recent work on nonclassicality~\cite{Bohmann2020}, which serves as a precursor to entanglement~\cite{Kim2002}, where the authors unify the notions of quasiprobability distributions and moment matrices. In this context, nonclassicality imprints on correlations between different phase-space coordinates. 

Here, we demonstrate that entanglement manifests in the partial derivatives of phase-space distributions. Given that any analytic function is described by its derivatives at a single point, this does not require full knowledge of the state's distribution. Instead, partial information in the form of a small subset of derivatives evaluated at the given coordinates is sufficient. We find that this approach significantly improves sensitivity to entanglement, a claim we corroborate by benchmarking the performance of the lowest-order criterion across important families of entangled states where other conventional witnesses fail. Our witnesses are readily accessible experimentally, as they require a surprisingly simple setup comprising a phase shift, a beamsplitter operation, state displacement, and particle-number-resolving detectors. This allows direct estimation of both the distribution and its partial derivatives at a single point, thus avoiding probing dense regions.

\textit{The remainder of this work is structured as follows.} In Sec.~\ref{sec:preliminaries}, we provide some preliminary information on phase-space distributions and the positive partial transpose criterion that underpins our approach. We derive our criteria in Sec.~\ref{sec:entanglementwitnesses}, followed by extensive benchmarks (Sec.~\ref{sec:examplestates}) and their measurement scheme (Sec.~\ref{sec:detection}). We discuss our findings and future directions in Sec.~\ref{sec:conclusions}.

\textit{Notation.} In this paper, we set $\hbar=1$ and use bold font to denote quantum operators, e.g., $\boldsymbol{x}$.

\section{Preliminaries}\label{sec:preliminaries}

\subsection{Phase-space distributions}
\label{subsec:PhaseSpaceDistributions}
We consider continuous-variable (CV) systems, characterised by infinite-dimensional Hilbert spaces and continuous, unbounded eigenspectra. In the case of a single mode, the state can be described by means of bosonic mode operators $\boldsymbol{a}$ and $\boldsymbol{a}^\dagger$ with $[\boldsymbol{a},\boldsymbol{a}^\dagger]=\mathds{1}$. These are related to the Hermitian position and momentum operators, $\boldsymbol{x}$ and $\boldsymbol{p}$, respectively, via
\begin{equation}
    \boldsymbol{a}=\frac{\boldsymbol{x}+\I\boldsymbol{p}}{\sqrt{2}} \quad \textrm{and} \quad \boldsymbol{a}^\dagger=\frac{\boldsymbol{x}-\I\boldsymbol{p}}{\sqrt{2}},
\end{equation}
with canonical commutation relations $[\boldsymbol{x},\boldsymbol{p}]=\I\mathds{1}$ understood. This generalises straightforwardly to $N$-mode systems~\cite{Schleich2001,Klauder2006,Mandel2013}.

Phase-space distributions are widely employed in the continuous-variable setting, as they provide a more convenient description of the state $\boldsymbol{\rho}$ under consideration. They are also commonly known as quasiprobability distributions, as they behave like probability distributions but violate Kolmogorov's $\sigma$-additivity axiom and may take on negative values. The most commonly used representations belong to the set of $s$-parametrized distributions~\cite{Cahill1969,Cahill1969_2,Agarwal1970}. Given a bipartite system consisting of two modes, the value of the distribution at the phase-space coordinate $(\alpha,\beta)\in\mathbb{C}^2$ may be expressed as
\begin{equation}\label{eq:P}\begin{split}
    &P(\alpha,\beta;\vec{\sigma})\\
    &=\left\langle:\frac{\sigma^{(a)}\sigma^{(b)}}{\pi^2}\exp{\left[-\sigma^{(a)}\boldsymbol{n}_a(\alpha)-\sigma^{(b)}\boldsymbol{n}_b(\beta)\right]}:\right\rangle,
\end{split}\end{equation}
where $:\cdot:$ indicates normal-ordering of the mode operators, and $\boldsymbol{a}$ and $\boldsymbol{b}$ represent the two modes. Therein, $\boldsymbol{n}_a(\alpha)=(\boldsymbol{a}-\alpha)^\dagger(\boldsymbol{a}-\alpha)$ is the displaced excitation-number operator of the mode $\boldsymbol{a}$, and similarly for $\boldsymbol{n}_b(\beta)$. For notational convenience, we introduce the width parameter $\vec{\sigma}=(\sigma^{(a)},\sigma^{(b)})$, where $\sigma^{(i)}\geq 0$ is the width parameter for mode $i$. This determines the extent to which the distribution is localised, or peaked, and is related to the parameter $s$ via 
\begin{equation}
    \sigma=2/(1-s).
\end{equation}
We note that this family of phase-space distributions is real-valued, $P(\alpha,\beta;\vec{\sigma})=P(\alpha,\beta;\vec{\sigma})^*$, and normalized, $\int\mathrm{d}^{4}\vec{\alpha}\,P(\alpha,\beta;\vec{\sigma})=1$. From this family, we identify the following commonly used distributions: the Husimi distribution $Q(\alpha,\beta)=P(\alpha,\beta;1,1)$ where $\sigma^{(i)}=1$ ($s^{(i)}=-1)$ for all $i$; similarly, the Wigner distribution $W(\alpha,\beta)=P(\alpha,\beta;2,2)$ at $\sigma^{(i)}=2$ ($s^{(i)}=0)$; and the Glauber--Sudarshan distribution $P_\text{GS}(\alpha,\beta)=P(\alpha,\beta;\infty,\infty)$ at $\sigma^{(i)}=\infty$ ($s^{(i)}=1$). Note that we also allow for $\sigma^{(i)}$ to be distinct for different $i$, which would imply different distributions for the different modes.

We note that care is needed when considering distributions for certain classes of states. For example, distributions with $\sigma>2$ may not be uniquely defined~\cite{Sperling2016} or cease to exist for certain states, whilst some Glauber-Sudarshan distributions ($\sigma\rightarrow\infty$) can become highly singular~\cite{Cahill1969,Cahill1969_2}. Therefore, we exclusively consider $\sigma \in [0,2]$ in the following.

\subsection{Negative partial transpose}

A bipartite state $\boldsymbol{\rho}$ is said to be separable if it can be expressed as a convex mixture of factorizable states:
\begin{equation}
    \boldsymbol{\rho}=\sum_ip_i\,\boldsymbol{\rho}_i^{(a)}\otimes\boldsymbol{\rho}_i^{(b)},
\end{equation}
where $\{\boldsymbol{\rho}_i^{(a)}\}_i$ and $\{\boldsymbol{\rho}_i^{(b)}\}_i$ are states of the first and second subsystems, respectively, and $\sum_ip_i=1$ with $p_i\geq0$ for all $i$. A necessary condition for separability is the Peres--Horodecki, or positive partial transpose (PPT), criterion~\cite{Peres1996,Horodecki1996}, whereby the state remains positive under partial transpose with respect to either of the subsystems,
\begin{equation}
\boldsymbol{\rho}^\textrm{PT}\sep 0,
\end{equation}
and, hence, a negative partial transpose of the state is a sufficient condition for entanglement.

In the case of CV systems, the partial transpose is equivalent to the time-reversal in one subsystem~\cite{Simon2000}. We consider the partial transpose with respect to mode $\boldsymbol{b}$, resulting in $\boldsymbol{b}\rightarrow\boldsymbol{b}^\dagger$. This led to the formulation of the Shchukin--Vogel hierarchy, derived in Ref.~\cite{Shchukin2005} (see also Refs.~\cite{Miranowicz2006,Miranowicz2009}). It is an infinite set of separability criteria based on moments of the mode operators and is equivalent to the PPT criterion. More precisely, a positive partial transpose is equivalent to
\begin{equation}\label{eq:ff0}
    \braket{\boldsymbol{f}^\dagger\boldsymbol{f}}^{\textrm{PT}}\geq 0
\end{equation}
for \emph{any} normal-ordered operator function $\boldsymbol{f}$. Here $\braket{\cdot}^\textrm{PT}$ denotes the expectation value with respect to $\boldsymbol{\rho}^\textrm{PT}$. After expanding $\boldsymbol{f}$ and invoking Sylvester's criterion, this is equivalent to the non-negativity of all principal minors of the matrix of moments:
\begin{equation}
    M=\begin{pmatrix}
        1 & \braket{\boldsymbol{a}} & \braket{\boldsymbol{a}^\dagger} & \braket{\boldsymbol{b}^\dagger} & \ldots \\
        \braket{\boldsymbol{a}^\dagger} & \braket{\boldsymbol{a}^\dagger\boldsymbol{a}} & \braket{\boldsymbol{a}^{\dagger 2}} & \braket{\boldsymbol{a}^\dagger\boldsymbol{b}^\dagger} & \ldots \\
        \braket{\boldsymbol{a}} & \braket{\boldsymbol{a}^2} & \braket{\boldsymbol{a}\boldsymbol{a}^{\dagger}} & \braket{\boldsymbol{a}\boldsymbol{b}^\dagger} & \ldots \\
        \braket{\boldsymbol{b}} & \braket{\boldsymbol{a}\boldsymbol{b}} & \braket{\boldsymbol{a}^\dagger\boldsymbol{b}} & \braket{\boldsymbol{b}^\dagger\boldsymbol{b}} & \ldots \\
        \vdots & \vdots & \vdots & \vdots & \ddots
    \end{pmatrix}\sep 0.
\end{equation}
Therefore, in order to detect entanglement, it is sufficient to show that at least one principal minor turns negative.

\section{Separability criteria}\label{sec:entanglementwitnesses}

We derive previously unknown separability criteria based on phase-space distributions and their derivatives. Their strongest form is equivalent to the PPT criterion. Unlike other criteria for CV systems, which generally rely on full quasiprobability distributions, their marginals, or their moments, this set of criteria depends only on the distribution's values at specific points in phase space. We discuss the simplest separability criterion of our hierarchy as well as multimode extensions.

\subsection{Infinite hierarchy of separability criteria}

We consider normally-ordered operators of the form
\begin{equation}\label{eq:fnp}\begin{split}
\boldsymbol{f}=\sum_{n,p}&c_{np}:\exp[-\sigma^{(a)}_{n}\boldsymbol{n}_a(\alpha)-\sigma^{(b)}_{p}\boldsymbol{n}_b(\beta)]:\\
&\qquad \qquad\times(\boldsymbol{a}-\alpha)^{n}(\boldsymbol{b}-\beta)^{p}.
\end{split}\end{equation}
with complex-valued coefficients $c_{np}$ and width parameters $\sigma_n^{(a)}, \sigma_{p}^{(b)} \in [0,2]$ comprising the Wigner representation ($\sigma=2$) and its smoothened variants ($0 \le \sigma < 2$). The expectation value of $\boldsymbol{f}^\dagger\boldsymbol{f}$ with respect to the partially transposed state has to satisfy Eq.~\eqref{eq:ff0} for all separable states, which translates into the bilinear form
\begin{equation}\label{eq:ffnpmq}\begin{split}
    &\braket{\boldsymbol{f}^\dagger \boldsymbol{f}}^\textrm{PT}=\sum_{n,p,m,q}c^*_{np}c_{mq}\,M_{np,mq} \sep 0,
\end{split}
\end{equation}
being non-negative for all $c_{np} \in \mathbb{C}$, or, equivalently, $M$ being positive semi-definite for all separable states.

After shifting the partial transpose from the state to the mode operators, the hermitian matrix $M$ reads
\begin{equation}\label{eq:Mnpmq0}\begin{split}
    M_{np,mq} & = \langle (\boldsymbol{a}-\alpha)^{\dagger n}(\boldsymbol{b}-\beta)^{\dagger q} \\
    &\hspace{0.5cm}:\exp[-\sigma^{(a)}_{nm}\boldsymbol{n}_a(\alpha)-\sigma^{(b)}_{pq}\boldsymbol{n}_b(\beta)]: \\
    &\hspace{0.6cm}(\boldsymbol{a}-\alpha)^{m}(\boldsymbol{b}-\beta)^{p}\rangle,
\end{split}
\end{equation}
with $:\hspace{-0.1cm}\exp[-\sigma_n\boldsymbol{n}]\hspace{-0.1cm}:\,:\hspace{-0.1cm}\exp[-\sigma_m\boldsymbol{n}]\hspace{-0.1cm}:\,=\,:\hspace{-0.1cm}\exp[-\sigma_{nm}\boldsymbol{n}]\hspace{-0.1cm}:$ understood, which follows from expanding the exponentials, and defining $\sigma_{nm}=\sigma_n+\sigma_m-\sigma_n\sigma_m \in [0,2]$\footnote{Allowing $\sigma_n>2$ results in $\sigma_{nm} < 0$ when $\sigma_m \to 2$, which prevents mapping the exponential in Eq.~\eqref{eq:Mnpmq0} to phase space.}. Next, we convert the remaining mode-operator polynomials into derivatives of the exponential kernel. Derivatives with respect to the width parameter generate creation and annihilation operators at equal order, to wit
\newpage
\begin{equation}
\begin{split}
    \label{eq:PhaseSpaceIdentity1}
    &\partial_\sigma^m\braket{:\exp{[-\sigma\boldsymbol{n}(\alpha)]}:}\\
    &=(-1)^m\braket{(\boldsymbol{a}-\alpha)^{\dagger m}:\exp{[-\sigma\boldsymbol{n}(\alpha)]}:(\boldsymbol{a}-\alpha)^{m}}.
    \end{split}
\end{equation}
In contrast, taking derivatives with respect to phase-space amplitudes leads to
\begin{equation}
    \label{eq:PhaseSpaceIdentity2}
    \begin{split}
        &\partial^m_{\alpha}\braket{:\exp{[-\sigma\boldsymbol{n}(\alpha)]}:}\\
        &=\sigma^m\braket{:\exp{[-\sigma\boldsymbol{n}(\alpha)]}:(\boldsymbol{a}-\alpha)^{ m}},
    \end{split}
\end{equation}
analogously for $\partial^m_{\alpha^*}\braket{:\exp{[-\sigma\boldsymbol{n}(\alpha)]}:}$. Here, we follow standard quantum optics conventions: the derivative with respect to the amplitude $\alpha$ is defined as $\partial_\alpha=\frac{1}{2}(\partial_{\Re[\alpha]}+\I\partial_{\Im[\alpha]})$, such that $\partial_\alpha\alpha=0$ and $\partial_\alpha \alpha^*=1$, and similarly for $\partial_{\alpha^*}$. Using Eqs.~\eqref{eq:PhaseSpaceIdentity1} and~\eqref{eq:PhaseSpaceIdentity2} in~\eqref{eq:Mnpmq0} and substituting the exponential with the general two-mode phase-space representation $P(\alpha,\beta;\sigma^{(a)}_{nm},\sigma^{(b)}_{pq})$ [see Eq.~\eqref{eq:P}] results in a closed expression for $M$ in phase space
\begin{equation}
\label{eq:Mnpmq}
\begin{split}
    M_{np,mq} &=\pi^2 (-1)^{n\wedge m+p\wedge q}\,\partial^{n\wedge m}_{\sigma^{(a)}_{nm}}\,\partial^{p\wedge q}_{\sigma^{(b)}_{pq}}\,\partial^{|n-m|}_{\overline{\alpha}}\partial^{|p-q|}_{\overline{\beta}} \\
    &\quad\qquad \times\frac{P(\alpha,\beta;\sigma^{(a)}_{nm},\sigma^{(b)}_{pq})}{\left(\sigma^{(a)}_{nm}\right)^{\abs{n-m}+1}\left(\sigma^{(b)}_{pq}\right)^{\abs{p-q}+1}}.
\end{split}\end{equation}
For compactness, we employed the notation $\partial_{x_0} f(x_0,y) \equiv \partial_xf(x,y)|_{x=x_0}$ and defined $i\wedge j \equiv \min (i,j)$ together with
\begin{equation}
    \partial_{\overline{\alpha}}^{\abs{n-m}}=\begin{cases}
        \partial_\alpha^{\abs{n-m}} &\textrm{for }m>n. \\ 
        1 &\textrm{for } m=n, \\
        \partial_{\alpha^*}^{\abs{n-m}} &\textrm{for } m<n,
    \end{cases}
\end{equation}
(analogously for $\partial^{|p-q|}_{\overline{\beta}}$). We remark that spatial and $\sigma$-derivatives are related via
\begin{equation}
    -\sigma^2\partial_{\sigma}\frac{P(\alpha;\sigma)}{\sigma}=\left(\sigma+\partial_{\alpha}\partial_{\alpha^*}\right)\frac{P(\alpha;\sigma)}{\sigma},
\end{equation}
which generalises straightforwardly to the two-mode case.

In a final step, we explicitly write $M$ as a matrix by choosing an ordering over multi-indices $i=(n,p)$ and $j=(m,q)$ such that $i<j$ if $n+p<m+q$ or $n+p=m+q$ and $p<q$, and defining the two-mode width parameter vector $\vec{\sigma}_{ij}=(\sigma^{(a)}_{ij},\sigma^{(b)}_{ij})$, which can also be parametrized in terms of $\sigma^{(a)}_{nm},\sigma^{(b)}_{pq}$ considered above. Then, the separability criterion~\eqref{eq:ffnpmq} is equivalent to 

\begin{widetext}\begin{equation}\label{eq:Msep0}
   \raisebox{-47pt}{\includegraphics[width=0.7\linewidth]{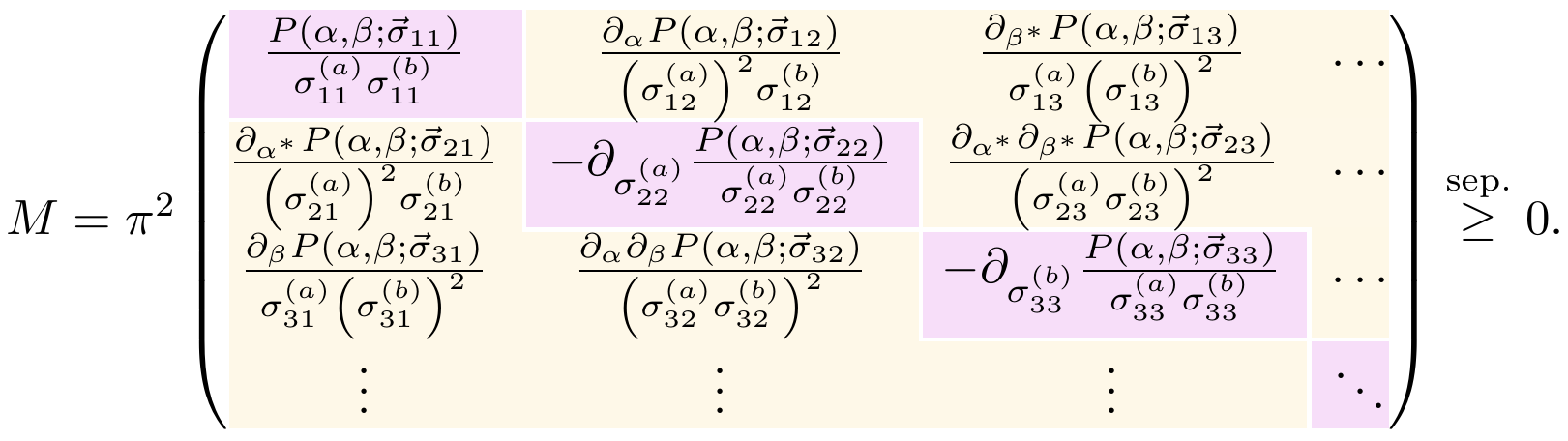}}
\end{equation}\end{widetext}
We obtain an infinite hierarchy of separability criteria in terms of phase-space distributions and their derivatives \emph{at specific points}. The structure of the matrix $M$ is graphically highlighted: the diagonal elements (purple) can be expressed solely in terms of $\sigma$-derivatives, whilst the off-diagonal elements (yellow) contain spatial derivatives. By invoking Sylvester's criterion, the matrix $M$ is positive-semidefinite if and only if all its principal minors are non-negative. Therefore, if any principal minor turns negative, entanglement is certified.

At this stage, we make a few remarks. First, we note that the phase-space coordinates $\alpha$ and $\beta$ are arbitrary for any given matrix, allowing us to optimize over phase space regions to achieve the strongest violation of a given criterion for a given state. Second, unlike in the case of nonclassicality~\cite{Bohmann2020}, $\alpha$ and $\beta$ are fixed for a given matrix and do not vary across the different elements. However, the two-mode width parameters summarized in $\vec{\sigma}_{ij}$ can be tuned across the different elements of the matrix. Third, first-order principal minors, i.e., diagonal elements of the matrix $M$, are invariant under partial transpose by Hermiticity of the matrix and, therefore, cannot be used on their own to detect entanglement.

\subsection{Equivalence with the PPT criterion}

We now show that~\eqref{eq:Msep0} is equivalent to the PPT criterion for the Husimi distribution ($\sigma=1$). By definition, a state $\boldsymbol{\rho}$ has a positive partial transpose if
\begin{equation}\label{eq:positivitydefn}
    \braket{\psi|\boldsymbol{\rho}^\textrm{PT}|\psi} = \textrm{Tr}[\rho^\textrm{PT}\boldsymbol{g}^\dagger\ket{00}\bra{00}\boldsymbol{g}] \geq 0
\end{equation}
holds for all state vectors $\ket{\psi}$. We use here that any state $\ket{\psi}$ can be expressed in terms of some operator-valued function acting on the two-mode vacuum, i.e., $\ket\psi=\boldsymbol{g}^\dagger\ket{00}$. We stress here that $\boldsymbol{g}^\dagger=g(\boldsymbol{a}^\dagger,\boldsymbol{b}^\dagger)$ can always be written as a power series in terms of creation operators only.

To relate the vacuum projector to the phase-space representations discussed above, we express the single-mode operator $:\exp[-\sigma\boldsymbol{n}]:$ in the Fock basis as
\begin{equation}\label{eq:expfockbasis}
    :\exp[-\sigma\boldsymbol{n}]:\;=\sum_{i=0}^{\infty} (1-\sigma)^i\ket{i}\bra{i}.
\end{equation}
This expression reflects the fact that all phase-space distributions with width parameters $\sigma\in[0,1]$ are non-negative. Generalising this to the two-mode case, the projector onto the two-mode vacuum state reads
\begin{equation}\begin{split}
    \ket{00}\bra{00} = &:\exp[-\sigma^{(a)}\boldsymbol{n}_a-\sigma^{(b)}\boldsymbol{n}_b]: \\
    & - \sum_{(i,j)\neq(0,0)}(1-\sigma^{(a)})^i(1-\sigma^{(b)})^j\ket{ij}\bra{ij},
\end{split}\end{equation}
which, when combined with~\eqref{eq:positivitydefn}, yields a reformulation of the PPT criterion
\begin{equation}\label{eq:fullinequality}\begin{split}
    & \braket{\boldsymbol{g}^\dagger:\exp[-\sigma^{(a)}\boldsymbol{n}_a-\sigma^{(b)}\boldsymbol{n}_b]:\boldsymbol{g} }^\textrm{PT}\\
    &\geq \sum_{(i,j)\neq(0,0)}(1-\sigma^{(a)})^i(1-\sigma^{(b)})^j\braket{ij|\boldsymbol{g}\boldsymbol{\rho}^\textrm{PT}\boldsymbol{g}^\dagger|ij},
\end{split}\end{equation}
for all $\boldsymbol{g}^\dagger=g(\boldsymbol{a}^\dagger,\boldsymbol{b}^\dagger)$. When restricting the width parameters to the range $\sigma^{(a)},\sigma^{(b)}\in[0,1]$, which includes the regime beyond the Husimi distribution ($\sigma=1$), the right hand side in~\eqref{eq:fullinequality} is lower bounded by zero, such that 
\begin{equation}\begin{split}\label{eq:simplerinequality}
    \braket{\boldsymbol{g}^\dagger:\exp[-\sigma^{(a)}\boldsymbol{n}_a-\sigma^{(b)}\boldsymbol{n}_b]:\boldsymbol{g}}^\textrm{PT}\sep 0,
\end{split}\end{equation}
for all $\boldsymbol{g}^\dagger=g(\boldsymbol{a}^\dagger,\boldsymbol{b}^\dagger)$. 

To relate~\eqref{eq:simplerinequality} with~\eqref{eq:ffnpmq}, which underlies our main result~\eqref{eq:Msep0} and probes entanglement across phase space, we introduce displacements. Indeed, the two-mode displacement operator is the unitary acting on the mode operators as $\boldsymbol{D}(\alpha,\beta)\,\boldsymbol{a}\boldsymbol{b}\,\boldsymbol{D}^\dagger(\alpha,\beta)=(\boldsymbol{a}-\alpha)(\boldsymbol{b}-\beta)$, with $\boldsymbol{D}(\alpha,\beta)^\dagger =\boldsymbol{D}(-\alpha,-\beta)$ understood. It is straightforward to check that non-negativity of the partially transposed state is conserved under arbitrary displacements, i.e., $\boldsymbol{\rho}^\textrm{PT}\geq 0 \iff \boldsymbol{D}(-\alpha,-\beta)\boldsymbol{\rho}^\textrm{PT}\boldsymbol{D}(\alpha,\beta)\geq 0$ for any $\alpha,\beta\in\mathbb{C}$. Hence, the separability criteria~\eqref{eq:simplerinequality} generalize to arbitrarily displaced operators
\begin{equation}\begin{split}\label{eq:simpledisplacement}
    \braket{\boldsymbol{D}(\alpha,\beta)\,\boldsymbol{g}^\dagger \hspace{-0.1cm}:\hspace{-0.05cm}\exp[-\sigma^{(a)}\boldsymbol{n}_a \hspace{-0.07cm}-\hspace{-0.05cm}\sigma^{(b)}\boldsymbol{n}_b] \hspace{-0.07cm}: \hspace{-0.07cm}\boldsymbol{g} \boldsymbol{D}(-\alpha,-\beta)}^\textrm{PT}
    \sep 0\, .
\end{split}\end{equation}
The above is equivalent to~\eqref{eq:ffnpmq}, and therefore also to~\eqref{eq:Msep0}, when setting $\sigma^{(a)}_{nm}=\sigma^{(a)}$ and $\sigma^{(b)}_{pq}=\sigma^{(b)}$ for all $n,m,p$ and $q$, in which case $\boldsymbol{f}^\dagger \boldsymbol{f} = \boldsymbol{D}(\alpha,\beta)\,\boldsymbol{g}^\dagger \hspace{-0.1cm}:\hspace{-0.05cm}\exp[-\sigma^{(a)}\boldsymbol{n}_a \hspace{-0.07cm}-\hspace{-0.05cm}\sigma^{(b)}\boldsymbol{n}_b] \hspace{-0.07cm}: \hspace{-0.07cm}\boldsymbol{g} \boldsymbol{D}(-\alpha,-\beta)$. At the same time,~\eqref{eq:simpledisplacement} is equivalent to the PPT criterion for $\sigma^{(a)}=1$ or $\sigma^{(b)}=1$ (which means selecting the Husimi distribution for at least one of the two modes), since the right-hand side of~\eqref{eq:fullinequality} vanishes in this case. Therefore, the positive semi-definiteness of the matrix $M$ in Eq.~\eqref{eq:Msep0} is equivalent to PPT when we set $\vec{\sigma}_{ij}=(1,1)$ for all $i,j$, i.e., when considering the Husimi distribution and its derivatives. 

On the other hand, the condition~\eqref{eq:simpledisplacement} is weaker than the PPT criterion for $0\leq \sigma^{(a)},\sigma^{(b)} <1$, see~\eqref{eq:fullinequality}. Note here that in setting $\sigma^{(a)}=\sigma^{(b)}=0$, one recovers moments-based criteria. Further, we remark that~\eqref{eq:simpledisplacement} does not apply for $\sigma^{(a)}$ or $\sigma^{(b)}$ greater than one, since the right-hand side of~\eqref{eq:fullinequality} is no longer necessarily non-negative for states with a positive partial transpose. \emph{This strongly suggests that we do not gain much in terms of entanglement sensitivity by considering distributions with $\sigma>1$.}

\subsection{Second-order separability criterion}

To gauge the detection capabilities of our approach, it is particularly interesting to consider the simplest second-order minor that captures cross-mode correlations and can, hence, detect entanglement. As we will show in Sec.~\ref{sec:detection}, this is also well suited for experimental implementation. In this section, we derive a specific class of criteria and formulate them in terms of the Husimi and Wigner distributions. We start by considering the lowest-order operator of the form~\eqref{eq:fnp} that contains cross-mode correlations, which reads
\begin{equation}\label{eq:f2}
\begin{split}
 \boldsymbol{f}&=c_1:\exp[-\sigma^{(a)}_{1}\boldsymbol{n}_a(\alpha)-\sigma^{(a)}_{1}\boldsymbol{n}_b(\beta)]:\\
 &\ \ +c_2:\exp[-\sigma^{(a)}_{2}\boldsymbol{n}_a(\alpha)-\sigma^{(b)}_{2}\boldsymbol{n}_b(\beta)]:(\boldsymbol{a}-\alpha)(\boldsymbol{b}-\beta).   
\end{split}
\end{equation}
Using~\eqref{eq:ffnpmq} we obtain the simple second-order criterion
\begin{align}\label{eq:som}
        M_2(\alpha,\beta;\vec{\sigma})&=\pi^2\left|\begin{matrix}
    \frac{P(\alpha,\beta;\vec{\sigma}_{11})}{\sigma^{(a)}_{11}\sigma^{(b)}_{11}} & \frac{\partial_{\alpha}\partial_{\beta^*}P(\alpha,\beta;\vec{\sigma}_{12})}{\left(\sigma^{(a)}_{12} \sigma^{(b)}_{12}\right)^2} \\
    \frac{\partial_{\alpha^*}\partial_{\beta}P(\alpha,\beta;\vec{\sigma}_{21})}{\left(\sigma^{(a)}_{21} \sigma^{(b)}_{21}\right)^2} & \partial_{\sigma^{(a)}_{22}}\partial_{\sigma^{(b)}_{22}}\frac{P(\alpha,\beta;\vec{\sigma}_{22})}{\sigma^{(a)}_{22}\sigma^{(b)}_{22}}
\end{matrix}\right| \nonumber \\
&\sep 0,
\end{align}
where again $\vec{\sigma}_{ij}=(\sigma^{(a)}_{ij},\sigma^{(b)}_{ij})$, with $\sigma^{(a)}_{ij}=\sigma^{(a)}_{i}+\sigma^{(a)}_{j}-\sigma^{(a)}_{i}\sigma^{(a)}_{j}\in[0,2]$ and similarly for $\sigma^{(b)}_{ij}$. Also, here we can choose all $\sigma^{(a)}_{i},\sigma^{(b)}_{i}$ to be distinct, which, together with the freedom in choosing the coordinate $(\alpha,\beta)$, results in a continuous family of separability conditions.

We consider two variants of the separability criteria in Eq.~\eqref{eq:som} to be of particular interest. When setting $\sigma_{i j}^{(a)} = \sigma_{i}^{(a)} = 1$ (analogously for $b$), we obtain criteria in terms of the Husimi distribution $Q(\alpha,\beta)$, to wit
\begin{equation} \label{eq:husimi}
    \begin{split}
        Q(\alpha,\beta)\left(1+\partial_{\alpha}\partial_{\alpha^*}\right)\left(1+\partial_{\beta}\partial_{\beta^*}\right)Q(\alpha,\beta)\\
        -\abs{\partial_\alpha\partial_{\beta^*} Q(\alpha,\beta)}^2 \sep 0.
    \end{split}   
\end{equation}

The inequality~\eqref{eq:som} may also be formulated in terms of the Wigner distribution, $W(\alpha,\beta)$, which is obtained by setting the width parameter to $\sigma=2$. Note that the Wigner distribution can not appear in the diagonal entries of $M_2$ since $\sigma^{(a)}_{i i} = \sigma^{(a)}_i (2 - \sigma^{(a)}_{i}) \in [0,1]$ when $\sigma^{(a)}_{i} \in [0,2]$ (similarly for $b$). However, we may set $\sigma^{(a)}_{1}=\sigma^{(b)}_{1}=0$ and $\sigma^{(a)}_{2}=\sigma^{(b)}_{2}=2$, thereby obtaining the criterion
\begin{equation}
\braket{\boldsymbol{n}_a(\alpha)\boldsymbol{n}_b(\beta)}-\frac{\pi^4}{16}\abs{\partial_\alpha\partial_{\beta^*} W(\alpha,\beta)}^2 \sep 0,
\end{equation}
where the first summand represents particle-number correlations under displacement.

\subsection{Generalizing to multimode systems}

Our phase-space approach generalises straightforwardly beyond two-mode systems. In the case of an $N$-mode system with $N>2$, one may check for entanglement across all possible bipartitions, and the PPT criterion has already been generalized to such settings \cite{Shchukin2006}. The phase-space distribution becomes
\begin{equation}
    P(\vec{\alpha};\vec{\sigma})=\prod_{i=1}^N\left\langle:\frac{\sigma^{(i)}}{\pi}\exp[-\sigma^{(i)}\boldsymbol{n}_i(\alpha_i)]:\right\rangle,
\end{equation}
where $\boldsymbol{n}_i$ is the excitation-number operator of the $i^\textrm{th}$ mode, and $\vec{\alpha}=(\alpha_1,\ldots,\alpha_N)$ and $\vec{\sigma}=(\sigma^{(1)},\ldots,\sigma^{(N)})$ are the vectors denoting the phase-space coordinate in $\mathbb{R}^{2N}$ and the $\sigma$-parameters.

We define the normally-ordered operator $\boldsymbol{f}=f(\boldsymbol{a}_1,\ldots,\boldsymbol{a}_N)$ similar to Eq.~\eqref{eq:fnp} as
\begin{equation}
    \boldsymbol{f} = \sum_{k_1,\ldots,k_N} c_{k_1\ldots k_N}\prod_{i=1}^N:\exp[-\sigma^{(i)}_{k_i}\boldsymbol{n}_i(\alpha_i)]:(\boldsymbol{a}_i-\alpha_i)^{k_i}.
\end{equation}
Then, we consider a bipartition across the modes, denoted by the complementary index sets $I \subset \{1, \dots, N \}$ and $\bar{I} = \{1, \dots, N \} \setminus I$. If the state is separable with respect to this bipartition, we have that $\braket{\boldsymbol{f}^\dagger\boldsymbol{f}}^{\textrm{PT},I}\sep 0$ for all such possible $\boldsymbol{f}$, where $\braket{\cdot}^{\textrm{PT},I}$ denotes the partial transpose with respect to the modes $i\in I$.

Using the same approach outlined earlier, we construct a matrix whose elements comprise phase-space distributions and their partial derivatives, and which must remain non-negative for all separable states. As an example, the second-order minor in Eq.~\eqref{eq:som} generalizes to
\begin{align}
    M_2&=\pi^N\left|\begin{matrix}
        \frac{P(\vec{\alpha};\vec{\sigma})}{\prod_i\sigma^{(i)}_{11}} & \frac{\prod_{i \in I, \bar{i}\in \bar{I}}\partial_{\alpha_{i}}\partial_{\alpha^*_{{\bar{i}}}} P(\vec{\alpha};\vec{\sigma})}{\prod_i \left(\sigma^{(i)}_{12} \right)^2} \\
        \frac{\prod_{i \in I, \bar{i}\in \bar{I}}\partial_{\alpha^*_{i}}\partial_{\alpha_{{\bar{i}}}} P(\vec{\alpha};\vec{\sigma})}{\prod_i \left(\sigma^{(i)}_{21} \right)^2} & (-1)^N\prod_i\partial_{\sigma^{(i)}_{22}}\frac{P(\vec{\alpha};\vec{\sigma})}{\prod_j \sigma^{(j)}_{22}}
    \end{matrix}\right| \nonumber \\
    &\sep 0.
\end{align}

\section{Example states}\label{sec:examplestates}
We demonstrate the strength and versatility of the phase-space approach for entanglement witnessing, with a focus on the prospects of our second-order criterion. In the following, we use the criterion defined in Eq.~\eqref{eq:som} with $\sigma \equiv \sigma^{(a)}_i = \sigma^{(b)}_i$ for $i=1, 2$, which we denote as $M_2(\alpha,\beta;\sigma)$.

\subsection{NOON states}

\begin{figure*}
    \centering
    \includegraphics[width=1.0\linewidth]{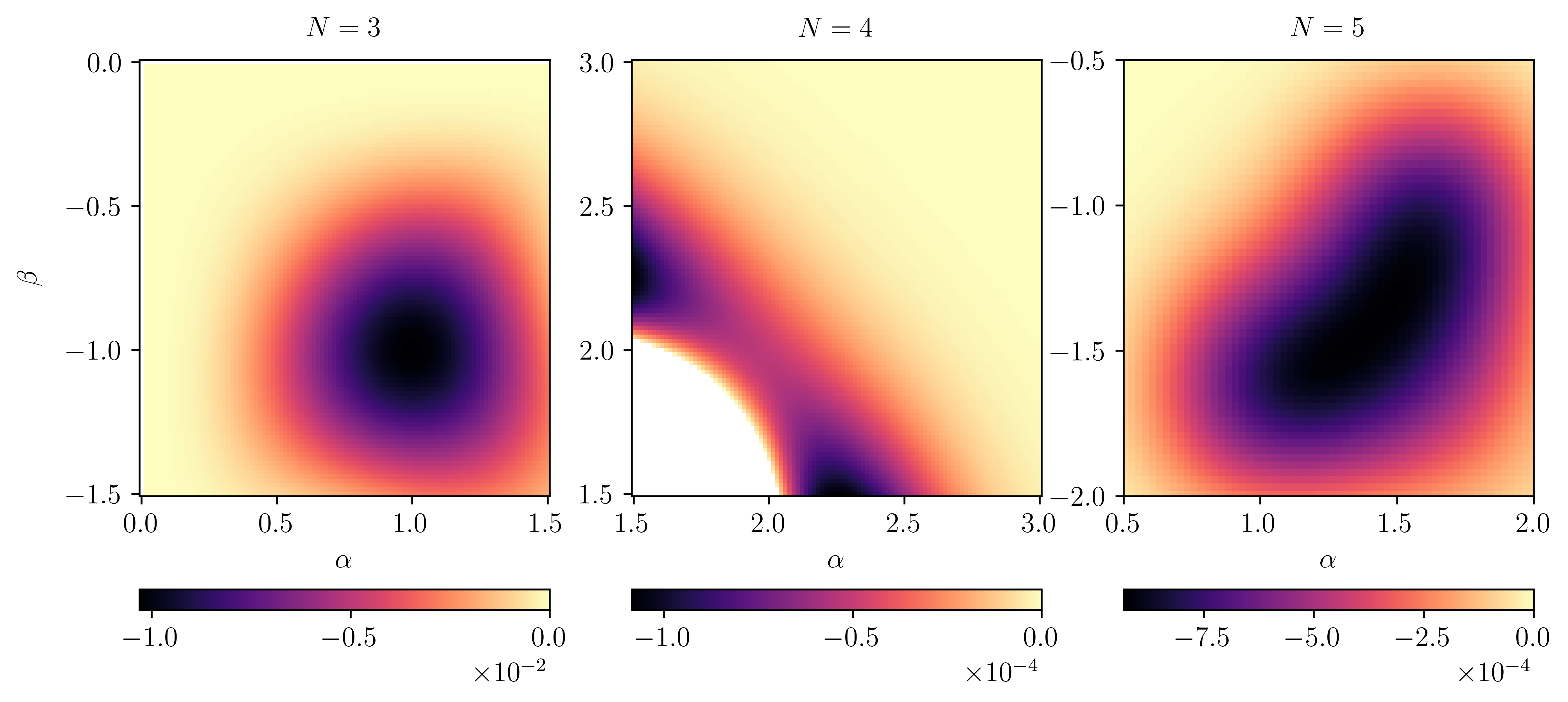}
    \caption{Negative regions of the minor $M_2(\alpha,\beta;1)$ for $N=3,4,5$ from left to right, with $\alpha,\beta$ being real. Entanglement is witnessed over entire regions rather than at isolated points in phase space, making the witness robust.}
    \label{fig:husiminoon}
\end{figure*}

\begin{figure}
    \centering
    \includegraphics[width=0.9\linewidth]{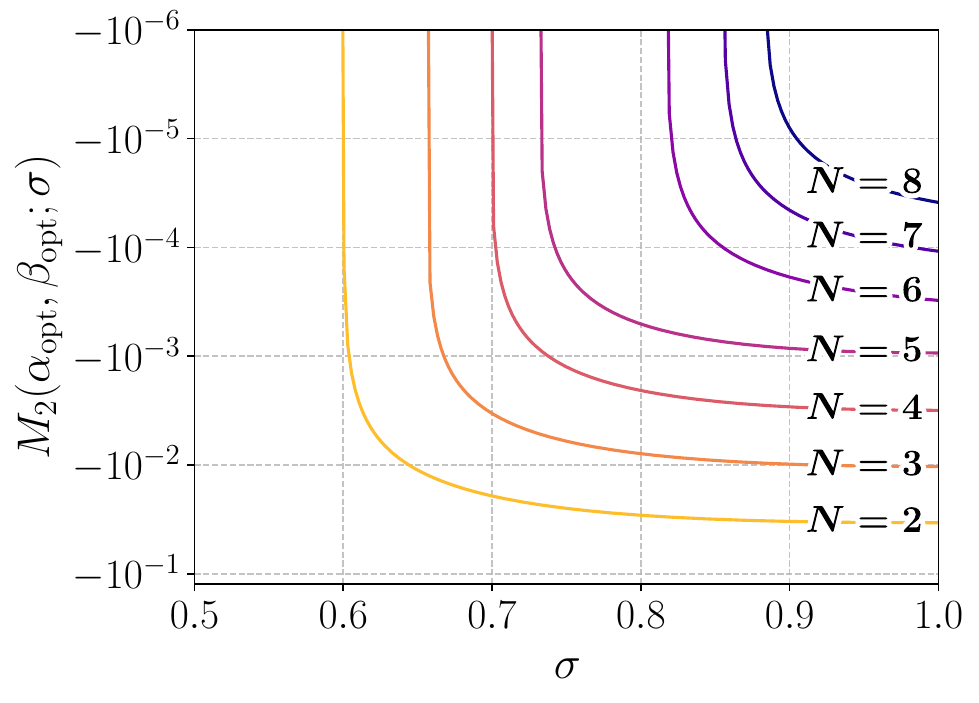}
    \caption{Separability criterion $M_2(\alpha_\textrm{opt},\beta_\textrm{opt};\sigma)\sep 0$ for various $N$ as a function of the $\sigma$-parameter, where $\alpha_\textrm{opt},\beta_\textrm{opt}$ are the coordinates for the global minimum of the Husimi-based criterion, $M_2(\alpha,\beta;1)$. The Husimi-based criterion is optimal, with increasing $N$ requiring larger $\sigma$ for flagging entanglement.}
    \label{fig:NOONsigma}
\end{figure}

\begin{figure}
    \centering
    \includegraphics[width=0.9\linewidth]{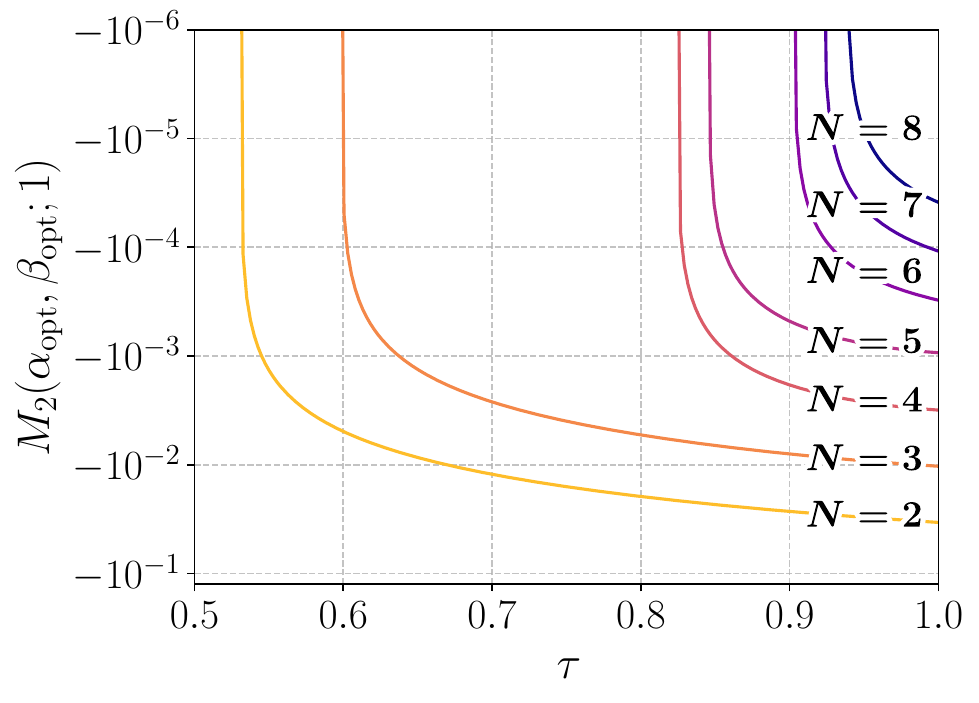}
    \caption{Separability criterion $M_2(\alpha_\textrm{opt},\beta_\textrm{opt};1)\sep 0$ for lossy NOON states~\eqref{eq:LossyNOONStates} with loss parameter $(1-\tau)$. Losses impact larger values of $N \gtrsim 6$, whilst entanglement is certified even for high loss-levels when $N \lesssim 3$.}
    \label{fig:NOONnoise}
\end{figure}

An important class of entangled states is the set of NOON states~\cite{Sanders1989}. They are relevant for applications such as quantum metrology~\cite{Dowling2008,Giovannetti2011,Huang2024} and are given by
\begin{equation}
    \ket{\Psi_\text{NOON}}=c_1\ket{N,0}+c_2\ket{0,N},
    \label{eq:NOONStates}
\end{equation}
where $\abs{c_1}^2+|c_2|^2=1$. These states are entangled for all $N \ge 1$ when $c_1,c_2\neq 0$. Yet, certifying their entanglement becomes increasingly difficult as $N$ increases: mode-operator-moment-based criteria stemming from the Shchukin--Vogel hierarchy require estimating moments of order $2N$~\cite{Callus2025}, which is experimentally difficult. Criteria based on partial-transpose-moments witness entanglement for all $N$, but presume three state copies~\cite{Deside2025}, which can become unfavourable as NOON states are often prepared probabilistically. No mixed-state criterion based on EPR-type variables, whether invoking second-order quadrature moments~\cite{Duan2000,Mancini2002,Giovannetti2003}, entropies~\cite{Walborn2009,Walborn2011,Saboia2011,Tasca2013,Schneeloch2018,Haas2022a}, or majorization relations~\cite{Haas2023a,Haas2023b}, detects entanglement. Only the stronger pure-state versions of the entropic criteria~\cite{Walborn2009,Saboia2011} certify entanglement for small $N$, while the entropic pure-state criterion in~\cite{Haas2021b} works for arbitrary $N$.

Focusing on the balanced NOON state with $c_1=c_2=1/\sqrt{2}$, we find that entanglement is certified by the second-order Husimi criteria~\eqref{eq:husimi} for arbitrary $N$. Starting from the Husimi distribution of the NOON state
\begin{equation}
    Q_\text{NOON}(\alpha,\beta)=\frac{1}{2N!}\abs{\alpha^N+\beta^N}^2 e^{-\abs{\alpha}^2-\abs{\beta}^2},
\end{equation}
we find that the inequality~\eqref{eq:husimi} is violated over various regions of phase space, see Fig.~\ref{fig:husiminoon}.

Further, whilst the Husimi-based criterion is an instance of the minor $M_2(\alpha,\beta;\sigma)$ in Eq.~\eqref{eq:som} with the $\sigma$-parameter set to 1, we find that the minor certifies entanglement over a wide range of $\sigma$. In Fig.~\ref{fig:NOONsigma}, we show that the Husimi-based criterion is optimal, while $\sigma$ can be tuned so that the criterion remains sensitive to entanglement for various excitation numbers $N$, with larger $N$ requiring distributions with larger $\sigma$.

To probe loss-robustness, we apply a loss channel to~\eqref{eq:NOONStates}. This amounts to mixing the NOON states with a two-mode vacuum state on a beamsplitter with transmittivity $\tau$~\cite{Bohmann2017,Deside2025}, which results in
\begin{equation}
    \begin{split}
        \boldsymbol{\rho}_\textrm{NOON}&=\frac{1}{2}\sum_{k=0}^N {N\choose k}\Big[\tau^{N-k}(1-\tau)^k \ket{N-k,0}\bra{N-k,0}\\
        &+\tau^k (1-\tau)^{N-k}\ket{0,k}\bra{0,k}\Big]\\
        &+\frac{\tau^N}{2}(\ket{N,0}\bra{0,N}+\ket{0,N}\bra{N,0}),
    \end{split}
    \label{eq:LossyNOONStates}
\end{equation}
such that $(1-\tau)$ is the loss parameter and $\tau=1$ represents the lossless case. In Fig.~\ref{fig:NOONnoise}, we show how such losses affect the sensitivity of the Husimi-based criterion $M_2(\alpha,\beta;1)$. We find that the witness is robust for smaller values of $N$, with larger $N$ resulting in more rapid deterioration with increasing losses.

\subsection{Entangled cat states}

\begin{figure}
    \centering
    \includegraphics[width=1.0\linewidth]{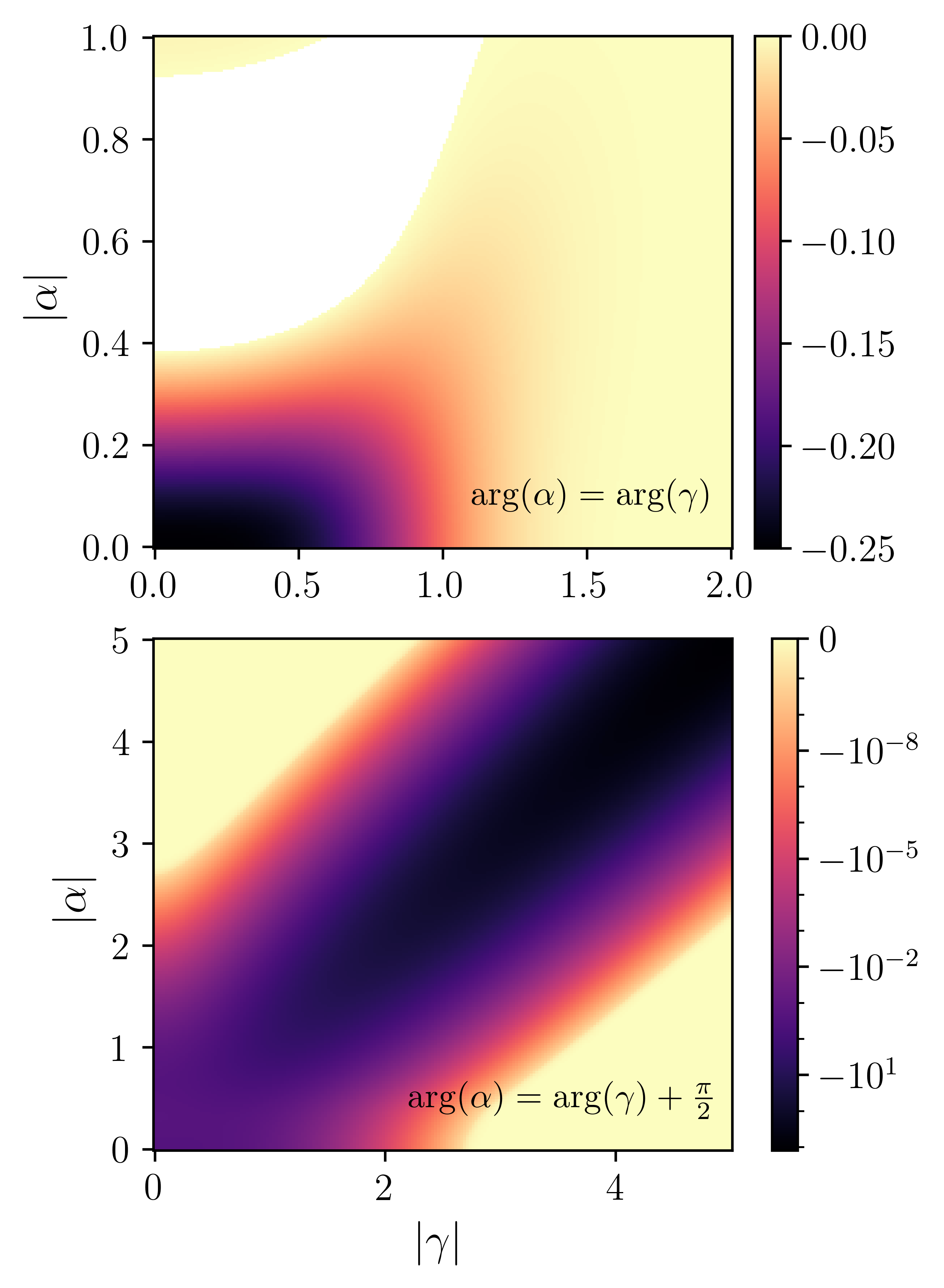} 
    \caption{Negative regions of the Husimi-based criterion $M_2(\alpha,\alpha;1)$ for the pure odd cat state $\boldsymbol{\rho}_\textrm{cat}(\gamma,\gamma,0)$. We evaluate the minor at $\arg(\alpha)=\arg(\gamma)$ (top, linear scale) and $\arg(\alpha)=\arg(\gamma)+\pi/2$ (bottom, log scale). The latter enables efficient entanglement detection for arbitrary coherent amplitudes $\abs{\gamma}$.}
    \label{fig:cattheta090}
\end{figure}

\begin{figure}
    \centering
    \includegraphics[width=1.0\linewidth]{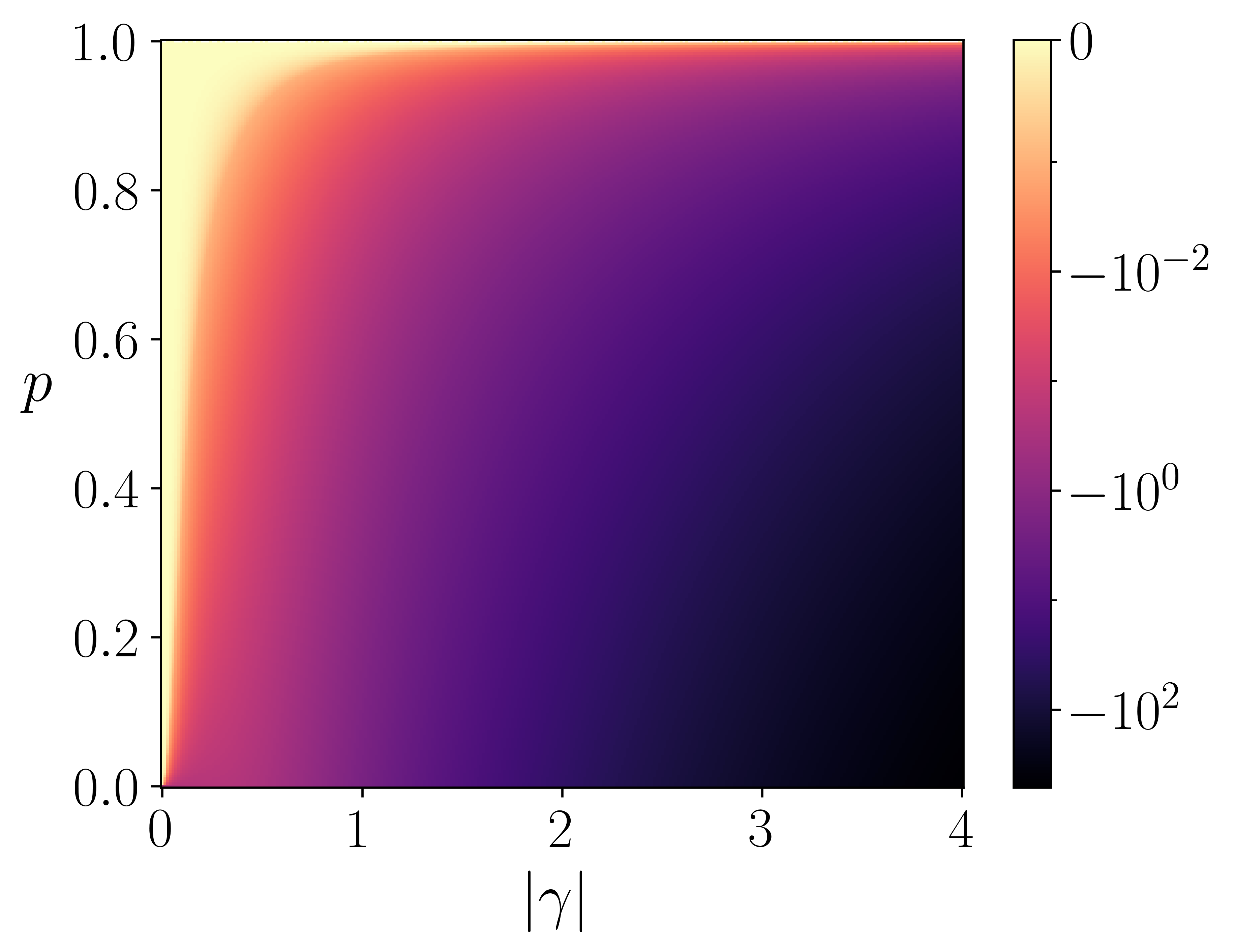}
    \caption{ $M_2(\I\gamma,\I\gamma;1)$ for odd cat states $\boldsymbol{\rho}_\textrm{cat}(\gamma,\gamma,p)$ for varying dephasing parameter $p$. Entanglement is certified for all $p<1$.}
    \label{fig:catdephasing}
\end{figure}

\begin{figure}
    \centering
    \includegraphics[width=1.0\linewidth]{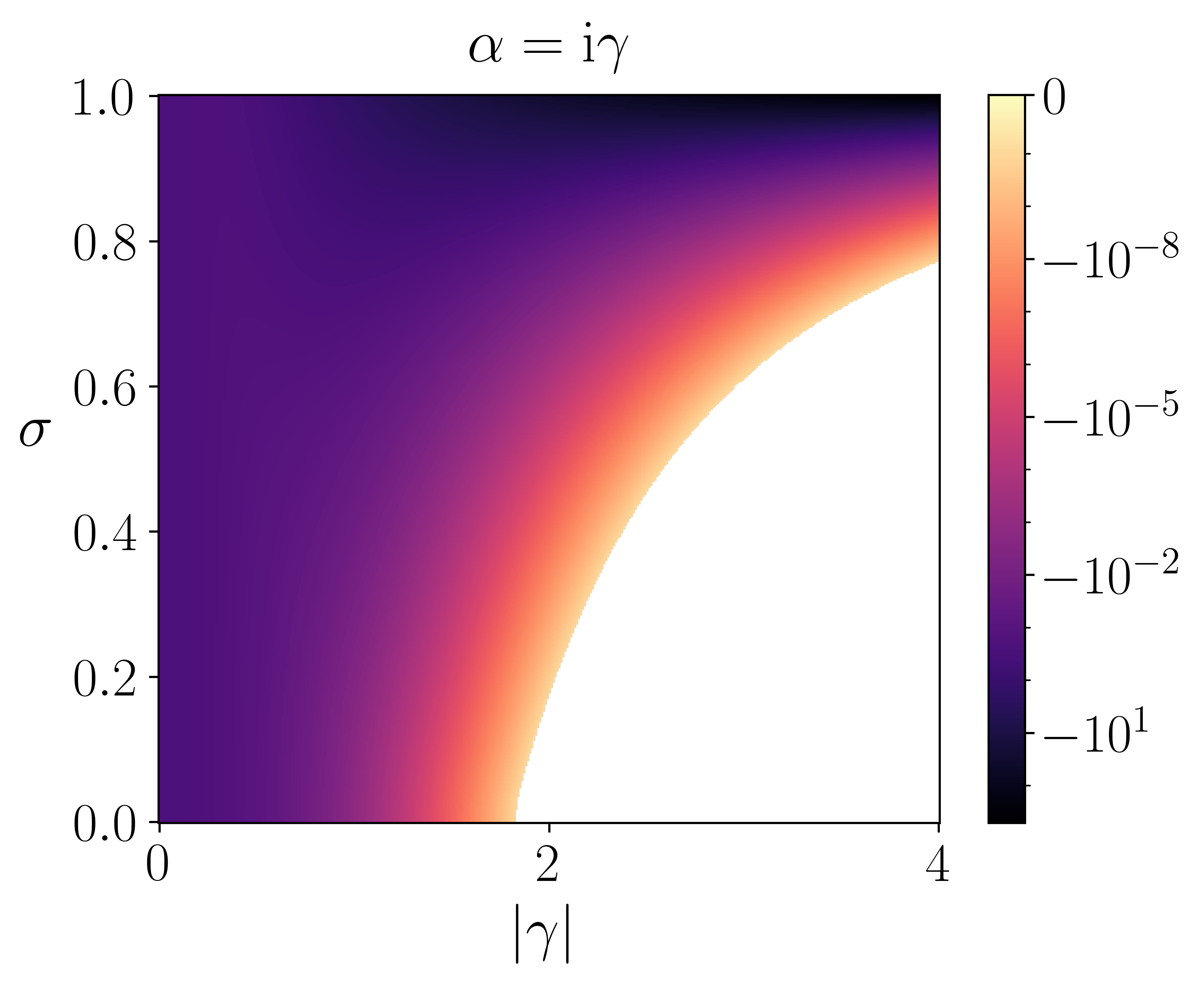}
    \caption{Regions where $M_2(\I\gamma,\I\gamma;\sigma) \ge 0$ is violated for pure odd cat states as a function of the width parameter $\sigma$. Increasing $\sigma$ improves the performance of the witness.}
    \label{fig:catsigma}
\end{figure}

Next, we consider the general class of entangled Schr\"{o}dinger cat states under dephasing, to wit
\begin{equation}\label{eq:catstate}\begin{split}
    \boldsymbol{\rho}_\text{cat}(\gamma,\delta,p)=\mathcal{N}(\gamma,\delta,p)\big[\ket{\gamma,\delta}\bra{\gamma,\delta}+\ket{-\gamma,-\delta}\bra{-\gamma,-\delta}\\
    +(1-p)(e^{\I\theta}\ket{\gamma,\delta}\ket{-\gamma,-\delta}+e^{-\I\theta}\ket{-\gamma,-\delta}\ket{\gamma,\delta}) \big],
\end{split}\end{equation}
where $\ket{\pm\gamma,\pm\delta}$ denote two-mode coherent states, $p\in[0,1]$ is the dephasing parameter which renders the state mixed for $p>0$ and pure otherwise, and $\mathcal{N}(\gamma,\delta)=1/[2+2(1-p)\cos\theta e^{-2(|\gamma|^2+|\delta|^2)}]$ specifies the normalisation constant. The phase $\theta \in [0, \pi]$ allows for smooth interpolation between even and odd cat states; we consider the latter by setting $\theta=\pi$ as the entanglement of even cat states is substantially simpler to detect~\cite{Callus2025,Deside2025}.

The state is entangled for all complex-valued $\gamma,\delta\neq 0$ and $p<1$. However, certifying entanglement for odd-cat states is highly nontrivial. Criteria based on mode-operator moments perform well for small coherent amplitudes $\abs{\gamma},\abs{\delta}\lesssim 2$ but suffer from exponentially small sensitivities beyond~\cite{Griffet2023b,Callus2025}. On the other hand, entropic criteria witness entanglement for large coherent amplitudes $\abs{\gamma},\abs{\delta}\gtrsim 3/2$~\cite{Walborn2009,Saboia2011,Haas2022a}.

In Fig.~\ref{fig:cattheta090} we show the Husimi-based criterion $M_2(\alpha,\beta;1)$ for pure cat states $\boldsymbol{\rho}(\gamma,\gamma,0)$. For $\arg(\alpha)=\arg(\gamma)$ (top), the witness performs well up until $\abs{\gamma} \lesssim 3/2$, similar to moment-based criteria. Remarkably, when evaluating the minor instead at $\arg(\alpha)=\arg(\gamma)+\pi/2$ (bottom), we find that entanglement is certified for arbitrary $\abs{\gamma}$. This demonstrates that our phase-space approach allows us to selectively probe regions of phase space where the interference and entanglement of coherent states manifest. Further, we find that the witness is also robust to dephasing, as shown in Fig.~\ref{fig:catdephasing}, with entanglement being witnessed until the state is fully dephased and, hence, separable. Finally, in Fig.~\ref{fig:catsigma}, we show that the separability criterion $M_2(\I\gamma,\I\gamma;\sigma)\sep 0$ is sensitive to odd-cat-state entanglement for all possible values of $\sigma$. More precisely, entanglement is verified for larger $\gamma$ with increasing $\sigma$, with $\gamma \to \infty$ when $\sigma \to 1$; hence, the Husimi witness is optimal once again.

\subsection{Random pure states}

\begin{figure}
    \centering
    \includegraphics[width=1.0\linewidth]{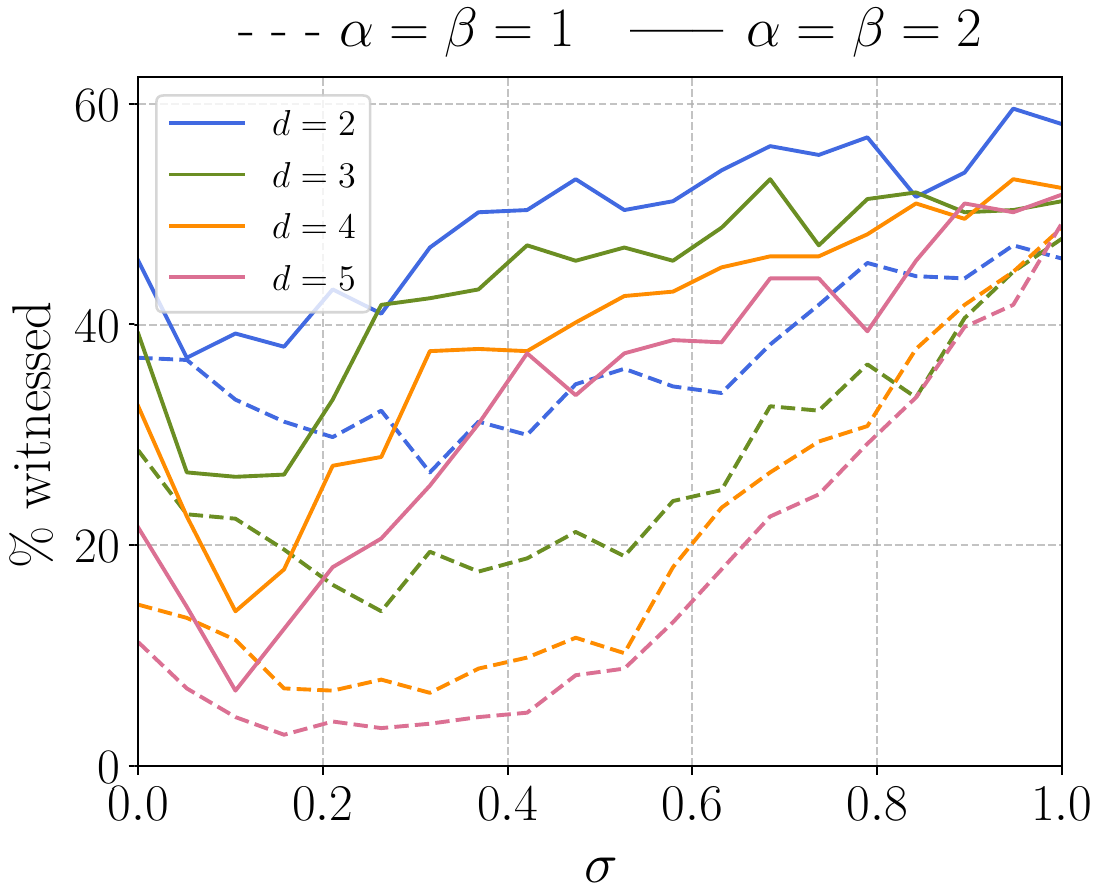}
    \caption{Rate at which entanglement of random pure states is witnessed using separability criterion $M_2(\alpha,\beta;\sigma)\geq 0$, where we set $\alpha=\beta=1$ (dashed lines) and $\alpha=\beta=2$ (solid lines). We sample from 500 randomly chosen states for each value of $\sigma$. Fluctuations indicate statistical errors (the curves smooth out when increasing sample size and fine-graining the discretization of $\sigma$).}
    \label{fig:randomnumber}
\end{figure}

Lastly, we consider pure states of the form
\begin{equation}
    \ket{\Psi_\textrm{rand}}=\sum_{i,j=0}^{d-1}c_{ij}\ket{ij},
\end{equation}
with random coefficients $c_{ij}$. The fraction of separable states quickly becomes vanishingly small as the dimension $d$ grows. In Fig.~\ref{fig:randomnumber}, we show how often the criterion~\eqref{eq:som} is violated for 500 Haar-randomly chosen states at different values of the width parameter $\sigma$. 

We first note that the witness fails when evaluating the minor at the origin, i.e., at $\alpha=\beta=0$. When considering $\alpha=\beta=1$ (dashed lines) and $\alpha=\beta=2$ (solid lines), we find that a significant fraction of the entangled states is detected, with an improvement for larger displacements $\alpha=\beta$ and a moderate decrease in the dimension $d$. This can be attributed to the spatial extent of phase-space distributions, which increases with the Fock number and requires probing the witness at larger displacements. We did not optimize the displacement amplitudes per state; hence, we expect better performance when minimizing over phase-space regions. Similarly, for small $d$, we expect slightly better performance when post-selecting on states that are actually entangled. As for the other considered example states, the witness is more effective for larger values of $\sigma$, with approx. $50\%$ of states witnessed when employing the Husimi distribution.

When compared with pre-existing criteria, third-order criteria based on partial-transpose moments automatically detect entanglement for all pure states~\cite{Deside2025}. However, accessing these witnesses requires three state replicas, whereas our second-order criteria can be accessed without interfering copies, as discussed in Sec.~\ref{sec:detection}. Entropy-based criteria valid for mixed states, such as those explored in~\cite{Walborn2009,Walborn2011}, are able to verify entanglement at a rate of up to $17.3\%$ for $d=2$ and $0.5\%$ for $d=3$, but require sampling the complete marginal distributions. In contrast, our criteria are based on the values of phase-space distributions and their derivatives at a single point in phase space.

\section{Detection scheme}\label{sec:detection}

\begin{figure}
    \centering
    \includegraphics[width=1.0\linewidth]{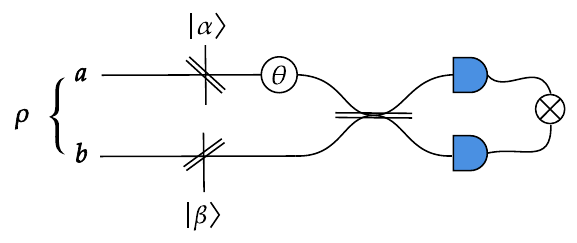}
    \caption{A schematic of a quantum optical setup that allows for the estimation of the second-order criterion for arbitrary $\sigma$-parameters, $M_2(\alpha,\beta;\vec{\sigma})$. The general setup consists of displacement operations via interference with coherent states $\ket{\alpha} $ and $ \ket{\beta}$ on separate beamsplitters, a phase shift $\theta$, a beamsplitter, and photon-number-resolving detectors at the two output arms.}
    \label{fig:schematic}
\end{figure}

We now present an approach to estimating the elements of the phase-space matrix underlying $M_2 (\alpha, \beta; \vec{\sigma})$ [see Eq.\eqref{eq:som}]. The main advantage of our approach is that we need to probe phase-space distributions and their derivatives, at an \emph{individual point} in phase space. We therefore do not need access to a fully reconstructed distribution. One obvious method for estimating derivatives, in particular low-order ones, at specific phase-space points is to evaluate the distribution's values in a small neighbourhood and estimate the derivatives from these values. Remarkably, however, all matrix elements can be accessed using a fairly simple measurement routine, which we sketch in Fig.~\ref{fig:schematic} following quantum optics conventions. We stress that all required operations have been demonstrated across various platforms. Given its superior performance, we first discuss our method for the Husimi-based entanglement witness, $M_2(\alpha,\beta;1)$, and generalize to arbitrary $\sigma$-parameters thereafter.

\subsection{Husimi-based criterion}

We start with the Husimi distribution of a state. Measuring this distribution has been long-established for quantum optical architectures~\cite{Noh1991,Noh1992,Leonhardt1993}, and can be realized in trapped-ion setups~\cite{Gaerttner2017}, Bose--Einstein condensates~\cite{Kunkel2022}, atoms in optical cavities~\cite{Haas2014,Barontini2015}, and circuit QED platforms~\cite{Kirchmair2013} (see~\cite{Lvovsky2009} for an overview). First, we consider the matrix element $M_2[1,1]$. Here, the general strategy is first to displace the state and then measure the projection onto the vacuum state $\ket{00}$~\cite{Lvovsky2009}. Indeed, we have that
\begin{equation}\label{eq:M211Husimi}\begin{split}
    M_2[1,1]=\pi^2Q(\alpha,\beta)&=\pi^2P(\alpha,\beta;1)=\braket{00|\boldsymbol{\rho}(\alpha,\beta)|00},
\end{split}\end{equation}
where we used the identity $:\exp(-\boldsymbol{n}_a-\boldsymbol{n}_b):=\ket{00}\bra{00}$. Here, $\boldsymbol{\rho}(\alpha,\beta)=\boldsymbol{D}^{\dagger}(\alpha,\beta)\boldsymbol{\rho}\boldsymbol{D}(\alpha,\beta)$ denotes a displaced state, with the two-mode displacement operator acting on the modes as $\boldsymbol{D}^\dagger(\alpha,\beta)\,\boldsymbol{a}\boldsymbol{b}\,\boldsymbol{D}(\alpha,\beta)=(\boldsymbol{a}+\alpha)(\boldsymbol{b}+\beta)$. The displacement of any quantum state can be implemented by interfering the state with a coherent state on a beam splitter with high transmittivity~\cite{Paris1996}.

Next, we consider the matrix element $M_2[2,2]$, for which we need to apply $\sigma$-derivatives. We find
\begin{equation}\begin{split}
   &\partial^i_{\sigma^{(a)}}\partial^j_{\sigma^{(b)}}\frac{\pi^2P(\alpha,\beta;\vec\sigma)}{\sigma^{(a)}\sigma^{(b)}}\bigg|_{\vec\sigma=(1,1)}\\
   &=(-1)^{i+j}\braket{\boldsymbol{D}(\alpha,\beta)\boldsymbol{a}^{\dagger i} \boldsymbol{b}^{\dagger j} \ket{00}\bra{00} \boldsymbol{a}^i \boldsymbol{b}^j \boldsymbol{D}^\dagger(\alpha,\beta)}\\
   &=(-1)^{i+j}\,i!\,j!\braket{ij|\boldsymbol{\rho}(\alpha,\beta)|ij}.
\end{split}\end{equation}
Here, the state is also displaced, but then projected onto the Fock state $\ket{ij}=\boldsymbol{a}^{\dagger i}\boldsymbol{b}^{\dagger j}\ket{00}/\sqrt{i!\,j!}$ instead, where $i$ and $j$ specify the orders of the derivatives for the first and second subsystem, respectively. For the second-order criterion, this reduces to $i=j=1$ and we obtain
\begin{equation}
    M_2[2,2]=\braket{11|\boldsymbol{\rho}(\alpha,\beta)|11}.
\end{equation}

Finally, we consider the matrix elements $M_2[1,2]$ and $M_2[2,1]$, which make up the last summand of our second-order criterion in Eq.~\eqref{eq:som}. The latter can be expressed more conveniently as
\begin{equation}\begin{split}\label{eq:measoff}
    &M_2[1,2]\times M_2[2,1]\\
    &=\left| \partial_\alpha\partial_{\beta^*}Q(\alpha,\beta)\right|^2=\frac{1}{\pi^4}\left|\braket{00|\boldsymbol{a}\,\boldsymbol{\rho}(\alpha,\beta)\,\boldsymbol{b}^\dagger|00}\right|^2.
\end{split}\end{equation}
We apply a balanced beam splitter with adjustable phase $\theta$ on mode $\boldsymbol{a}$, acting as
\begin{equation}\label{eq:beamsplitter}
    \begin{split}
        &\boldsymbol{a}\rightarrow\boldsymbol{c}_{\theta}=(e^{\I\theta}\boldsymbol{a}+\boldsymbol{b})/\sqrt{2},\\
        &\boldsymbol{b}\rightarrow\boldsymbol{d}_{\theta}=(e^{\I\theta}\boldsymbol{a}-\boldsymbol{b})/\sqrt{2},
    \end{split}
\end{equation}
on the displaced state. When measuring the projection onto the Fock states $\ket{10}$ at $\theta =0$ and $\theta=\pi/2$, we find:
\begin{subequations}
    \begin{equation}\begin{split}
        \theta=0: \quad&\braket{00|\boldsymbol{c}_{0}\,\boldsymbol{\rho}(\alpha,\beta)\,\boldsymbol{c}_0^\dagger|00}\\
        &=\braket{00|\boldsymbol{a}\,\boldsymbol{\rho}(\alpha,\beta)\,\boldsymbol{a}^\dagger|00}+\braket{00|\boldsymbol{b}\,\boldsymbol{\rho}(\alpha,\beta)\,\boldsymbol{b}^\dagger|00}\\
        &\quad+2\Re\left[\braket{00|\boldsymbol{a}\,\boldsymbol{\rho}(\alpha,\beta)\,\boldsymbol{b}^\dagger|00}\right],
    \end{split}\end{equation}
\begin{equation}\begin{split}
        \theta=\frac{\pi}{2}: \quad & \braket{00|\boldsymbol{c}_{\pi/2}\,\boldsymbol{\rho}(\alpha,\beta)\,\boldsymbol{c}_{\pi/2}^\dagger|00}\\
        &=\braket{00|\boldsymbol{a}\,\boldsymbol{\rho}(\alpha,\beta)\,\boldsymbol{a}^\dagger|00}+\braket{00|\boldsymbol{b}\,\boldsymbol{\rho}(\alpha,\beta)\,\boldsymbol{b}^\dagger|00}\\
        &\quad-2\Im\left[\braket{00|\boldsymbol{a}\,\boldsymbol{\rho}(\alpha,\beta)\,\boldsymbol{b}^\dagger|00}\right],
    \end{split}\end{equation}
\end{subequations}
where $\Re[\cdot]$ and $\Im[\cdot]$ denote the real and imaginary parts, respectively. The term in Eq.~\eqref{eq:measoff} can then be accessed by independently estimating $\braket{00|\boldsymbol{a}\,\boldsymbol{\rho}(\alpha,\beta)\,\boldsymbol{a}^\dagger|00}$ and $\braket{00|\boldsymbol{b}\,\boldsymbol{\rho}(\alpha,\beta)\,\boldsymbol{b}^\dagger|00}$ via setting the beam splitter transmittivity to zero and unity, respectively, which corresponds to measuring the projection of the original state onto $\ket{10}$ and $\ket{01}$. The Husimi-based criterion is therefore favourable not only because it is optimal for the states we have considered, but also because it requires the estimation of only the zero- and single-photon projections.

\subsection{Arbitrary $\sigma$-parameter}

Next, we generalize our scheme to access the second-order criterion $M_2(\alpha,\beta;\vec{\sigma})$ in Eq.~\eqref{eq:som} to any choice of width parameters with $\sigma^{(a)}_{ij}=\sigma^{(b)}_{ij}\equiv\sigma_{ij} \in [0,1]$ for $i,j=1,2$, based on approaches described in Refs.~\cite{Wallentowitz1996,Banaszek1999,Shchukin2005b,Bohmann2020}. The setup is the same as for the Husimi-based criterion, shown in Fig.~\ref{fig:schematic}, and the initial state $\boldsymbol{\rho}$ is to be displaced in the same manner, $\boldsymbol\rho \rightarrow \boldsymbol{\rho}(\alpha,\beta)$. The difference lies in the need to introduce a generalized measurement. Following Eq.~\eqref{eq:expfockbasis}, the required operator reads
\begin{equation}
    \boldsymbol{E}_{11}=\sum_{i,j}(1-\sigma_{11})^i(1-\sigma_{11})^j \ket{ij}\bra{ij},
\end{equation}
which contains negative elements when $\sigma_{11} > 1$. This is expected, as phase-space representations exhibit negative values in this regime. Intuitively speaking, $\boldsymbol{E}_{11}$ smoothly interpolates between estimating the Wigner distribution via parity measurements, for which $\boldsymbol{E}_{11} = \sum_{i j} (-1)^{i + j} \ket{ij} \bra{ij}$, and the vacuum projection resulting in the Husimi distribution, in which case $\boldsymbol{E}_{11} = \ket{0 0} \bra{0 0}$, see Eq.~\eqref{eq:M211Husimi}. Then, measuring the general matrix element $M_2[1,1]$ in Eq.~\eqref{eq:som} amounts to
\begin{equation}
    M_2[1,1]=\frac{\pi^2P(\alpha,\beta;\vec{\sigma}_{11})}{\sigma_{11}^2}=\textrm{Tr}[\boldsymbol{E}_{11} \boldsymbol{\rho}(\alpha,\beta)].
\end{equation}

Similarly, the matrix element $M_2[2,2]$ is determined using the measurement operator
\begin{equation}\begin{split}
    \boldsymbol{E}_{22}&=\sum_{i,j\geq 0}(1-\sigma_{22})^i(1-\sigma_{22})^j \boldsymbol{a}^\dagger \boldsymbol{b}^\dagger\ket{ij}\bra{ij}\boldsymbol{a} \boldsymbol{b}\\
    &= \sum_{i,j\geq 0}(1-\sigma_{22})^i(1-\sigma_{22})^j (i+1)(j+1)\\
    &\qquad \qquad \times\ket{i+1,j+1}\bra{i+1, j+1}
    \end{split}\end{equation}
such that
\begin{equation}
    \begin{split}
    M_2[2,2] &=\partial_{\sigma^{(a)}}\partial_{\sigma^{(b)}}\frac{\pi^2P(\alpha,\beta;\vec{\sigma})}{\sigma^{(a)}\sigma^{(b)}}\bigg|_{\sigma^{(a)},\sigma^{(b)}=\sigma_{22}} \\
    &=\textrm{Tr}[\boldsymbol{E}_{22} \boldsymbol{\rho}(\alpha,\beta)].
\end{split}\end{equation}

Finally, to obtain the partial derivatives $\partial_\alpha \partial_{\beta^*}$ and their complex conjugates, we introduce a beamsplitter interaction in the same way as for the Husimi-based criterion, see Eq.~\eqref{eq:beamsplitter}. Note that, by restricting to $\sigma_{ij,1}=\sigma_{ij,2}$, the action of the beamsplitter on the measure $\boldsymbol{E}_{11}$ is simply
\begin{equation}\begin{split}
    \boldsymbol{E}_{11}^{(\boldsymbol{a},\boldsymbol{b})}=&\sum_{i,j}(1-\sigma_{12})^i(1-\sigma_{12})^j \ket{ij}\bra{ij}_{\boldsymbol{a},\boldsymbol{b}}\\
    \xrightarrow[]{\textrm{beamsplitter}}&\sum_{i,j}(1-\sigma_{12})^i(1-\sigma_{12})^j \ket{ij}\bra{ij}_{\boldsymbol{c},\boldsymbol{d}}=\boldsymbol{E}_{11}^{(\boldsymbol{c},\boldsymbol{d})},
\end{split}\end{equation}
where $\boldsymbol{E}_{11}^{(\boldsymbol{a},\boldsymbol{b})}$ denotes the representation of the measurement operator in terms of the mode operators $\boldsymbol{a}$ and $\boldsymbol{b}$, and similarly for $\boldsymbol{E}_{11}^{(\boldsymbol{c},\boldsymbol{d})}$. Then, we obtain the following probabilities when setting $\theta=0$ and $\theta=\pi/2$:
\begin{subequations}
    \begin{equation}\begin{split}
        \theta=0: \;\; & \textrm{Tr}[\boldsymbol{c}_0^\dagger\boldsymbol{E}_{11}^{(\boldsymbol{c},\boldsymbol{d})}\boldsymbol{c}_0 \boldsymbol{\rho}(\alpha,\beta)] \\
&= \textrm{Tr}[\boldsymbol{a}^\dagger\boldsymbol{E}_{11}^{(\boldsymbol{a},\boldsymbol{b})}\boldsymbol{a}\, \boldsymbol{\rho}(\alpha,\beta)] + \textrm{Tr}[\boldsymbol{b}^\dagger\boldsymbol{E}_{11}^{(\boldsymbol{a},\boldsymbol{b})}\boldsymbol{b}\, \boldsymbol{\rho}(\alpha,\beta)]\\
& \quad + 2\Re\left[\textrm{Tr}[\boldsymbol{b}^\dagger\boldsymbol{E}_{11}^{(\boldsymbol{a},\boldsymbol{b})}\boldsymbol{a} \boldsymbol{\rho}(\alpha,\beta)]\right],
    \end{split}\end{equation}
\begin{equation}\begin{split}
\theta=\frac{\pi}{2}: \;\; & \textrm{Tr}[\boldsymbol{c}_{\pi/2}^\dagger\boldsymbol{E}_{11}^{(\boldsymbol{c},\boldsymbol{d})}\boldsymbol{c}_{\pi/2} \boldsymbol{\rho}(\alpha,\beta)] \\
&= \textrm{Tr}[\boldsymbol{a}^\dagger\boldsymbol{E}_{11}^{(\boldsymbol{a},\boldsymbol{b})}\boldsymbol{a}\, \boldsymbol{\rho}(\alpha,\beta)] + \textrm{Tr}[\boldsymbol{b}^\dagger\boldsymbol{E}_{11}^{(\boldsymbol{a},\boldsymbol{b})}\boldsymbol{b}\, \boldsymbol{\rho}(\alpha,\beta)]\\
& \quad - 2\Im\left[\textrm{Tr}[\boldsymbol{b}^\dagger\boldsymbol{E}_{11}^{(\boldsymbol{a},\boldsymbol{b})}\boldsymbol{a} \boldsymbol{\rho}(\alpha,\beta)]\right],
    \end{split}\end{equation}
\end{subequations}
where the matrix element $M_2[1,2]$ is
\begin{equation}\begin{split}
    M_2[1,2] &=\frac{\pi^2\partial_\alpha \partial_{\beta^*} P(\alpha,\beta;\vec{\sigma}_{12})}{\sigma_{12}^2} \\
    &=\textrm{Tr}[\boldsymbol{b}^\dagger\boldsymbol{E}_{11}^{(\boldsymbol{a},\boldsymbol{b})}\boldsymbol{a} \boldsymbol{\rho}(\alpha,\beta)],
\end{split}\end{equation}
and $M_2[2,1]=\overline{M_2[1,2]}$. 

Importantly, our scheme offers the advantage of choosing suitable width parameters $\vec{\sigma}_{ij}$ in post-processing: the optical setup allows for the establishment of the joint photon-number distribution, whilst the weighting of the probability distribution, which corresponds to different $\sigma$-parameters, can be tuned post-measurement.

\section{Discussion \& outlook}\label{sec:conclusions}

In this work, we presented an infinite hierarchy of separability criteria that rely solely on local features of $\sigma$-parametrised phase-space distributions. This overcomes the difficulty in, say, entropic separability criteria, which require estimating full (quasi-)probability distributions, and provides a new angle on the moments-based approach. We have shown that the Husimi-based variant of our approach is equivalent to the PPT criterion. Although our criteria complement the Shchukin--Vogel hierarchy, which is based on moments of the mode operators, we demonstrated that our phase-space approach enhances entanglement-detection capabilities across various classes of states. Indeed, we focused on the lowest-order minor and showed how it outperforms its moments-based analogues. We provided simple routines to access our family of criteria, thereby bypassing the need to estimate derivatives by evaluating distributions in a neighbourhood of the chosen phase-space coordinate. Indeed, our criteria can be experimentally accessed using a simple circuit implementable on any platform that supports state displacement, effective phase-shift, and beamsplitter operations, along with particle-number-resolving detection---including photonics~\cite{Slussarenko2019} and ultracold atoms~\cite{Viermann2022,Deller2025a,Deller2025b}. Our scheme is also advantageous because it does not require copies or ancilla measurements, unlike other existing protocols. Further, the Husimi-based criterion requires only resolution of the zero- and single-particle-number sectors.

We have observed how, within the set of phase-space-based witnesses, the Husimi-based criterion ($\sigma=1$) outperforms criteria where $\sigma<1$, with respect to various entangled states. We remark that violation of a separability condition for some $\sigma<1$ does not imply that the \emph{same} inequality is violated at the \emph{same} point in phase space when setting $\sigma=1$. Indeed, it is not too difficult to construct a counterexample. However, it would be interesting to investigate whether a refined argument holds: if a state violates the non-negativity of some minor for $\sigma<1$, does there always exist some point in phase space for which the minor with $\sigma=1$ turns negative as well? If true, the Husimi distribution would be optimal and the simplest to measure, as it requires only low-order Fock-state projections.

While we have shown the versatility of our second-order criterion, it would be interesting to investigate further why the lowest-order witness of our hierarchy suffices to detect the non-Gaussian entanglement of such a wide range of entangled states. This could be done by characterizing the set of states that do not violate the inequality after scanning all of phase space. Furthermore, it is highly relevant to consider sampling complexity and the effect of finite detector efficiency on the feasibility of entanglement certification using our approach. This could also be combined with a study of how the $\sigma$-parameter and the displacement amount can be optimized to overcome difficulties in sampling from photon-number distributions. Finally, an interesting future direction is to investigate how our phase-space approach could be generalized to, e.g., finite-dimensional quantum systems. Here, phase-space techniques include discrete distributions~\cite {Wootters1987} as well as continuous distributions on curved phase spaces, such as the 2-sphere for spins~\cite{Zhang1990}, for which estimating the Husimi distribution is well understood~\cite {Amiet1999}.

\begin{acknowledgments}
    This research is supported by funding from the German Research Foundation (DFG) under the project identifiers 398816777-SFB 1375 (NOA) and 550495627-FOR 5919 (MLCQS), from the Carl-Zeiss-Stiftung within the QPhoton Innovation Project MAGICQ, from the Federal Ministry of Research, Technology and Space (BMFTR) under project BeRyQC, by the BMBF project PhoQuant (Grant No. 13N16110), and the EU project SPINUS (Grant No. 101135699).
\end{acknowledgments}

\bibliography{bibliography}

\end{document}